\documentstyle[12pt,epsfig]{article}
\newcommand{\STRUT}{\rule{0in}{2.5ex}}
\begin{document}
\hfill{DOE/ER/40427-22-N95}

\hfill{LA-UR-95-4023-REV}

\begin{center}

{\Large {\bf Phenomenological Aspects of Isospin Violation\\
in the Nuclear Force}}\\

\vspace*{0.20in}
by\\
\vspace*{0.20in}
U.\ van Kolck \\
Department of Physics \\
University of Washington \\
Seattle, WA  98195 \\
\vspace*{0.20in}
and\\
\vspace*{0.20in}
J.L.\ Friar and T.\ Goldman \\
Theoretical Division \\
Los Alamos National Laboratory \\
Los Alamos, NM  87545 \\

\end{center}
\vspace*{0.25in}

\begin{abstract}

Phenomenological Lagrangians and dimensional power counting are used to assess
isospin violation in the nucleon-nucleon force.  The $\pi NN$ coupling constants
(including the Goldberger-Treiman discrepancy), charge-symmetry breaking, and
meson-mixing models are examined. A one-loop analysis of the isospin-violating
$\pi NN$ coupling constants is performed using chiral perturbation theory.
Meson-mixing models and the $^3$He - $^3$H mass difference are also discussed
in the context of naturalness.
\end{abstract}

\pagebreak

Isospin violation in the strong interaction remains one of the least understood
aspects of the nuclear force \cite{hm}.  After the long-range and
well-understood electromagnetic interaction is removed from the nucleon-nucleon
($NN$) interaction, small but significant differences ($\sim$ 1\%) exist between
the $nn$, $pp$, and (T = 1) $np$ interactions.  We briefly discuss these 
differences below in the context of phenomenological Lagrangians that exhibit
(broken) chiral symmetry as well as isospin violation \cite{vk1}.  A detailed
discussion will be presented elsewhere \cite{vk2}.  Our topics are:  (1) isospin
violation in the $\pi NN$ coupling constants; (2) the sizes of $\rho - \omega$
and $\pi - \eta$ mixing forces \cite{msc,cp}, as well as forces from 
$(h_1, b_1)$ and $(a_1, f_1)$ meson mixing; (3) naturalness and the 
$^3$He - $^3$H mass difference.

Isospin violation in the $NN$ force is usually described \cite{hm} in terms of
classes of possible isospin operators for nucleons 1 and 2.  These can be taken
to be: (I) 1 and $\mbox{\boldmath$t$} (1) \cdot \mbox{\boldmath$t$} (2)$; (II)
$t_z (1) t_z (2) - \mbox{\boldmath$t$} (1) \cdot \mbox{\boldmath$t$} (2)$; (III)
$t_z (1) + t_z (2)$; (IV) $[\mbox{\boldmath$t$} (1) \times \mbox{\boldmath$t$}
(2)]_z$ and $(t_z (1) - t_z (2))$, and are referred to as:  (I)
charge-independent; (II) charge-dependent; (III) and (IV)  charge asymmetric or
charge-symmetry breaking (CSB).  Classes (II) and (IV) forces vanish for $pp$ 
and $nn$ systems, while class (III) vanishes for the $np$ system.  Class (IV)
forces involve T = 0 and T = 1 mixing in the $np$ system.  The best evidence for
class (II) forces is obtained from $NN$ phase-shift analyses (PSA).  The recent
Nijmegen PSA \cite{ds1,ds2} finds a fairly strong difference between $np$ (T =
1) and $pp$ $^1{\rm S}_0$ phase shifts and determines them separately.  The
numbers that best characterize this charge dependence are the $^1{\rm S}_0$
scattering lengths after removal of all long-range electromagnetic effects:
$a_{np}$ ($-23.75$ fm) and $a_{pp}$ ($-17.3$ fm). This PSA also finds that the
$^3P_J$ waves are charge dependent.

The various $\pi NN$ coupling constants $f^2$ can be written in the generic form
$$
f^2 = \biggl( \frac{1}{4 \pi} \biggr) \biggl( \frac{g_{A} \, m_{\pi^{+}} \, d}
{2f_{\pi}} \biggr)^2 \, , \eqno (1)
$$
where $g_{A}$ is the axial vector coupling constant (1.2573(28)) \cite{pd},
$f_{\pi}$ is the pion decay constant (92.4(3) MeV) \cite{pd}, the charged pion
mass is conventional and is entered to make $f^2$ dimensionless, while $d-1$ is
the Goldberger-Treiman (GT) discrepancy \cite{gt}. The latter is positive and a
measure of chiral-symmetry breaking and is defined in terms of the pseudoscalar
form of the $\pi NN$ coupling constant, $G$, and the nucleon mass, $M$, by $G/M 
= g_A d/f_{\pi}$. Because the Nijmegen PSA explicitly includes the tail of the
$NN$ force (OPEP), this procedure can be used to determine all of the $\pi NN$
coupling constants, $\pi^0 pp$, $\pi^0nn$, and $\pi^{\pm} np$, as well as the
masses of the exchanged pions. Proton-proton scattering determines both the
$\pi^{0}$ mass and $f^2_p = f^2_{pp \pi^{0}}$, while neutron-proton scattering
determines $f^2_0 = f_{nn \pi^{0}} f_{pp \pi^{0}}$ and $f^2_c =  f_{np \pi^{-}}
f_{pn \pi^{+}}$ and the charged-pion mass. This PSA finds \cite{ds1} an
exchanged $\pi^0$ mass of 135.6(13) MeV and a $\pi^{\pm}$ mass of 139.4(10) MeV
and the $pp$ \cite{ds1} and $np$ \cite{ds2} results (separated by a bar) listed
in Table 1.

\begin{table}[hbt]
\centering
\caption{Pion-nucleon coupling constants determined by the 
Nijmegen\protect\cite{ds1,ds2} PSA.}

\begin{tabular}{|c|cc|}
\hline
$f^2_{\pi^0 pp}$ & $f_{\pi^0 pp} f_{\pi^0 nn}$ & $f^2_{\pi^c np}$ \STRUT \\ 
\hline \hline
0.0745(6) & 0.0745(9) & 0.0748(3) \STRUT\\ \hline
\end{tabular}
\end{table}

\noindent We will return to these results shortly. We further note that setting
$d$ to 1 in eqn.\ (1) produces $f^2 (d=1)$ = 0.0718(5), which demonstrates that
the measured values are close to the chiral limit.

A separate experiment \cite{pg} on $\pi^- + d \rightarrow n + n + \gamma$
determined the $nn$ $^1{\rm S}_0$ scattering length. After removal of small
electromagnetic effects \cite{hm} this is found to be $a_{nn} = -18.8(4)$ fm. 
The difference of $-1.5(5)$ fm between $a_{nn}$ and $a_{pp}$ is the best 
experimental information \cite{hm} that we have on CSB in the $NN$ force. Other
evidence comes from the $^3$He - $^3$H binding energy difference of 764 keV.
Approximately 693 keV is attributable to the Coulomb interaction between protons
(648 keV) and to small nucleon magnetic moment and velocity-dependent parts of
the Breit interaction and to the nucleons' mass difference in the kinetic energy
(45 keV) \cite{wu}.  The remaining $\sim$ 70(25) keV is consistent with any
short-range CSB interaction that produces the $-1.5(5)$ fm difference between
$nn$ and $pp$ scattering lengths.

A systematic development was recently made \cite{vk1} of isospin-violating
interactions in the context of the general effective chiral Lagrangian, which we
briefly summarize now.  These have three distinct origins: (1) the mass
difference of the $u$- and $d$-quarks in QCD specified by $\epsilon \equiv
\frac{m_{d} -m_{u}}{m_{d} + m_{u}} \sim$ 0.3 \cite{rabi}; (2) the (frozen-out)
effect of high-frequency photons exchanged between quarks in a nuclear system;
(3) the effect of soft photons that are exchanged or modify vertices. Because
only a subset of the last type of process has been implemented (or even
calculated) we will not further consider these corrections. They are expected to
be of order ($\alpha/\pi$) times the OPEP coupling constants and therefore
probably not larger than the errors on those quantities \cite{fc}.

This formalism is designed to separate the effects of the short-range QCD
dynamics from those of long range, which are mostly constrained by chiral
symmetry. The former are usually calculated using simple models, while in our
approach they appear in the Lagrangian as parameters whose values are not
determined by symmetry. We make only the minimal assumption of naturalness
\cite{'th}, namely, that unless suppressed by some symmetry these parameters are
given by naive dimensional analysis \cite{gm}. The (magnitudes of the)
dimensionless factors that result from such an analysis should be numbers that
are ${\cal O}(1)$ (we will see later that in our case all are in the range
0.5-2.0). The long-range effects are obtained from the pion cloud, including the
contributions of loops.

We wish to perform a one-loop calculation of isospin violation in the
one-pion-exchange potential (OPEP) due to the first two mechanisms listed above,
and later discuss shorter-range effects. We can organize the calculation using
power counting \cite{w} to characterize the individual vertices: $ \Delta = 
{\rm d} + n_F /2 -2 $, where d is the number of derivatives or pion masses at
each vertex and $n_F$ is the number of fermion fields, while $\Delta$ determines
the number of powers of a typical small momentum (i.e., $Q^{\Delta}$)
characteristic of an amplitude.  Chiral symmetry dictates that $\Delta$ must be
nonnegative for strong interactions. Adding a loop to an amplitude constructed
from these building blocks increases the exponent of $Q$ by 2. (Electromagnetic
interactions can produce vertices with smaller values of $\Delta$, but are
suppressed by powers of $\alpha (\sim 1/137)$.)

There are three relevant isospin-conserving interactions of lowest order 
($\Delta = 0$):
$$
L^{(0)}_{IC} = - \frac{g_A}{f_{\pi}} \; \overline{N} 
\mbox{\boldmath $\sigma$} \cdot \mbox{\boldmath $\nabla$} 
(\mbox{\boldmath $t$} \cdot \mbox{\boldmath  $\pi$}) N \,
(1 - \frac{\mbox{\boldmath  $\pi$}^2}{4 f^2_{\pi}})
-\frac{1}{2 f^2_{\pi}} \overline{N} 
\mbox{\boldmath $t$} \times \mbox{\boldmath  $\pi$} \cdot 
\dot{\mbox{\boldmath  $\pi$}} N + \frac{m^2_{\pi} \mbox{\boldmath  $\pi$}^4
-2 \mbox{\boldmath  $\pi$}^2 ({\partial_\mu \mbox{\boldmath  $\pi$}})^2}
{8 f^2_{\pi}} \, . \eqno (2a)
$$
There are three appropriate isospin-violating interactions arising from 
quark-mass differences ($L^{(\Delta)}_{qm}$),
$$
L_{qm} = -\delta M \; 
\overline{N} (t_3 - \frac{\pi_3 \mbox{\boldmath $t$} \cdot 
\mbox{\boldmath  $\pi$}}{2 f_{\pi}^2}) N 
+ \frac{\beta_1}{2 f_{\pi}} \; \overline{N} 
\mbox{\boldmath $\sigma$} \cdot \mbox{\boldmath $\nabla$} 
\pi_3 N +\frac{\delta M}{2 M^2} \; \overline{N} \mbox{\boldmath $p$}^2 t_3 N 
\, , \eqno (2b)
$$
corresponding to $\Delta$ = 1, 2, and 3, respectively. The first and third 
terms reflect differences in the nucleon rest and kinetic masses ($\delta M
\sim \epsilon m^2_{\pi}/\Lambda$), while $\beta_1 \sim \epsilon
m^2_{\pi}/\Lambda^2$ is an isospin-violating $\pi NN$ coupling constant. The
constants in eqn.\ (2b) are chiral-symmetry breaking and hence proportional to
$\epsilon m^2_{\pi}$, while $\Lambda$ is the QCD large-mass scale ($\sim$ 1
GeV). The powers of $m_{\pi}$ contribute to the power counting.

There are also isospin-violating interactions of the EM type 
($L^{(\Delta)}_{EM}$)
$$
L_{EM} = -\frac{\bar{\delta} m^2_{\pi}}{2} (\mbox{\boldmath  
$\pi$}^2 -\pi^2_3 )(1 - \frac{\mbox{\boldmath $\pi$}^2}{2 f^2_{\pi}})
- \bar{\delta} M \, \overline{N} t_3 N +
\frac{\bar{\beta}_3}{2 f_\pi} \overline{N} 
\mbox{\boldmath $\sigma$} \cdot \mbox{\boldmath $\nabla$} \pi_3 \, N 
$$
$$
+\frac{\bar{\beta}_{10}}{2 f_{\pi}}\, \overline{N} t_3 
\mbox{\boldmath $\sigma$} \cdot \mbox{\boldmath $\nabla$} \pi_3 \, N 
+ \left( \frac{\bar{\beta}_4 + \bar{\beta}_5}
{2 f_{\pi}}\right)\overline{N}  
\mbox{\boldmath $\sigma$} \cdot \mbox{\boldmath $\nabla$} (\mbox{\boldmath  
$\pi$} \times \mbox{\boldmath $t$})_3\, N \, , \eqno (2c)
$$
where $\bar{\delta} m^2_{\pi} \sim  \alpha \Lambda^2 /\pi$, $\bar{\delta} M \sim
\bar{\delta} m^2_{\pi}/ \Lambda$, and $\bar{\beta}_i \sim \bar{\delta}
m^2_{\pi}/ \Lambda^2$. The first two terms are of orders $-2$ and $-1$,
respectively, while the remaining three are of order 0. The last term in eqn.\
(2c) will not contribute to OPEP because of symmetry, while $\bar{\delta} M$ can
be added to $\delta M$ to form the complete nucleon mass and $\bar{\beta}_3 +
\beta_1$ is the complete CSB $\pi NN$ vertex, which we will continue to denote
by $\beta_1$.

The large pion-mass difference is primarily of electromagnetic origin, and this
allows the relative sizes of the interactions due to quark-mass differences
$(L^{(\Delta)}_{qm})$ and hard electromagnetic processes $(L^{(\Delta)}_{EM})$
to be related \cite{vk1,vk2}.  One expects on this basis that $L^{(-2)}_{EM}
\sim L^{(1)}_{qm}$, and the first, second, and remaining terms in eqn.\ (2c) can
be taken to be comparable to terms in eqn.\ (2b) of orders 1, 2, and 3. In order
to be consistent to this order we also need one-loop corrections (equivalent to
$\Delta$ = 3) obtained from a single $L^{(1)}_{qm}$ or $L^{(-2)}_{EM}$ and
various $L^{(0)}_{IC}$ terms from our Lagrangian
$$
\Delta L = L_{IC} + L_{qm} + L_{EM} + \cdot \cdot \cdot \cdot \cdot \, .
\eqno (2d)
$$
Only nonanalytic terms are kept, since the others can be absorbed
into the parameters of the Lagrangian (see, for example, ref.\cite{bss}).

\begin{figure}[htb]
  \epsfig{file=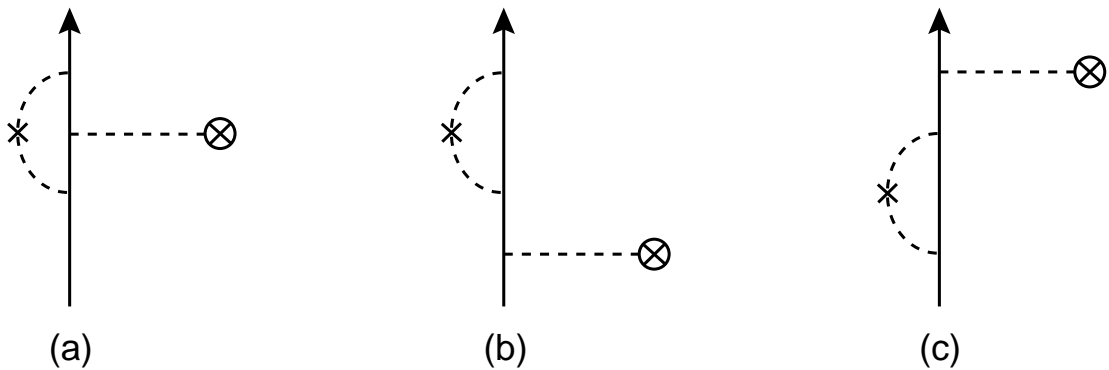,height=1.9in}
  \caption{The pion-mass-difference effect in a pion vertex correction is shown
in (a), while the same effect on the nucleon self energy is illustrated in
(b) and (c). Pions are indicated by dashed lines, nucleons by solid lines,
and the charged-neutral pion-mass difference is indicated by a cross. The
circled cross indicates an off-shell pion.}
\end{figure}

We find that 6 classes of diagrams have nonvanishing contributions to this 
order. Four classes cancel in pairs: isospin-violating nucleon self-energy
graphs containing nucleon-mass differences cancel the corresponding vertex
corrections, while the isospin-conserving pion loops containing a pion-mass
difference in vertex bubbles (from the first term in $L_{IC}$) cancel the
corresponding pion self-energy bubbles (from the last term in $L_{IC}$). The
graphs in Figure (1) remain. The nucleon self-energy graphs are isospin
conserving. The vertex correction in Fig.\ (1a) gives the sole isospin-violating
nonanalytic term
$$
\Delta \bar{\beta}_{10} =  \frac{2 \, \bar{\delta} m^2_{\pi}\, g^3_A \,
\log (\mu/m_{\pi})}{(4 \pi f_{\pi})^2} \, , \eqno (3)
$$
where $\mu$ is the renormalization scale. Choosing $\mu \sim m_{\rho}$ gives
$\Delta \bar{\beta}_{10} = 6 \cdot 10^{-3}$. Such terms are special because of
their analytic structure and because they are typically larger. Our results
below for $\bar{\beta}_{10}$ will include $\Delta \bar{\beta}_{10}$.

The one-pion-range force generated by the interactions in eqn.\ (2) has the 
form in momentum space
$$
V_{\pi}{(\mbox{\boldmath $q$})} = \left[ \frac{g_A}{f_{\pi}^2}  
\frac{\mbox{\boldmath  $\sigma$} (1) \cdot \mbox{\boldmath $q$} 
\mbox{\boldmath  $\sigma$} (2) \cdot \mbox{\boldmath $q$}}
{\mbox{\boldmath $q$}^2 + m^2_{\pi}}\right] \left( g_A \, d^2 \,
\mbox{\boldmath $t$}(1) \cdot \mbox{\boldmath $t$}(2) -\frac{\beta_1}{2} 
(t_z(1) + t_z(2)) -\bar{\beta}_{10} t_z (1) t_z (2) \right) \, ,
\eqno (4)
$$
and generates charge-asymmetric and charge-dependent $\pi NN$ coupling 
constants in the usual (charge-independent) OPEP. Because the latter is
conventionally defined in terms of $f^2$ (as in eqn.\ (1)), it involves $d$ and
is proportional to $\frac{g^2_A d^2}{f_{\pi}^2} \mbox{\boldmath $t$}_1 \cdot
\mbox{\boldmath $t$}_2$.  In the experimental data we identify an
isospin-conserving Goldberger-Treiman discrepancy, $d-1$, and attribute any
isospin violation to $\beta_1$ and $\bar{\beta}_{10}$. That is, we use eqn. (1)
to define and extract from the data the quantities $d_{pp}, d_{nn}, d_c$, and
$d_0 (\sim 1)$, from which $d, \beta_1$, and $\bar{\beta}_{10}$ can be easily
extracted:
$$
\beta_1 = \frac{g_A}{2} \left(d_{nn} - d_{pp}\right), \eqno (5a)
$$
$$
d = d_c \; , \eqno (5b)
$$
$$
\bar{\beta}_{10} = g_A (2 d_c -d_{pp} -d_{nn}) \; .\eqno (5c)
$$
If $pp$ data are not used we cannot determine $\beta_1$ and we must replace
eqn.\ (5c) by
$$
\bar{\beta}_{10} = 2 g_A (d_c - d_0)  \; . \eqno (5d)
$$
The results are shown in Table 2. 

\begin{table}[hbt]
\centering
\caption{GT discrepancy and isospin-violating $\pi NN$ coupling constants 
determined from the Nijmegen PSA.}

\begin{tabular}{|ccc|c|}
\hline
$d-1$ & $\beta_1$ & $\bar{\beta}_{10}$ & Type \STRUT \\ 
\hline \hline
2.1(5)\% & 0(9) $\cdot 10^{-3}$ & 5(18)$ \cdot 10^{-3}$ & $np$ and $pp$ \STRUT\\
2.1(5)\% &                      & 5(18)$ \cdot 10^{-3}$ & $np$ only \\ \hline
\end{tabular}

\end{table}

As noted earlier, the GT discrepancy is smaller than many previous values, but
is consistent with the dimensional estimate: $d-1 \sim m^2_{\pi}/\Lambda^2 \sim
2\%$. The isospin-violating parameters $\beta_1$ and $\bar{\beta}_{10}$ are
consistent with zero.  The parameter $\beta_1$ should be of order $(\epsilon
m^2_{\pi} / \Lambda^2 )$ or $\sim 6 \cdot 10^{-3}$, which is slightly smaller
than the uncertainty in $\beta_1$.  A direct determination of $\beta_1$ would
probably require reducing the uncertainty by a factor of two, and this would be
quite difficult. The nonanalytic contribution to $\bar{\beta}_{10}$ produced in
the one-loop calculation is of comparable size. Unfortunately the error bars are
twice as large in this case.

Note that the rest masses of the nucleons can play no role outside of loops,
while the different nuclear kinetic masses in eqn.\ (2b) are properly treated in
the Nijmegen PSA.

In addition to the one-pion-range potentials there are isospin-violating
short-range potentials. The leading-order terms have $\Delta$ = 2 and are
given by \cite{vk1,vk2}
$$
L^{(2)}_{qm} =  \gamma_{s} (\overline{N} t_3 N) 
(\overline{N} N) + \gamma_{\sigma} (\overline{N} t_3 \mbox{\boldmath 
$\sigma$} N) \cdot (\overline{N} \mbox{\boldmath $\sigma$} N) \, ,  \eqno (6)
$$
where $\gamma_i \sim \epsilon \, m^2_{\pi} /f^2_{\pi} \, \Lambda^2$. According
to the definitions presented earlier, these interactions will generate nuclear
forces of class (III). As expected from power counting, the pion-mass difference
from eqn.\ (2c) generates a relatively large and well-known contribution to
class (II) through differences in the range of OPEP that it produces. Class (IV)
forces, on the other hand, appear only in higher orders; as a consequence of the
structure of the chiral Lagrangian, they are therefore the smallest
\cite{vk1,vk2}.

We now briefly discuss mesonic models. In the well-studied mesonic sector it has
been found that the parameters of the chiral Lagrangian are reliably estimated
if they are assumed to be saturated by resonance exchange \cite{rs}. Analogous
models based on isospin mixing have been used extensively to estimate CSB in the
$NN$ force (see for example refs.\cite{msc,cp}). The isospin mixing of $\rho$
and $\omega$ mesons produces a class (III) force \cite{msc}, which for small
$\mbox{\boldmath $q$}^2 ( << m^2_{\omega} \sim m^2_{\rho})$ takes the
momentum-space form
$$
V_{\rho \omega}(\mbox{\boldmath $q$}) = 
\frac{- g_{\rho} g_{\omega} \langle \rho^0 | H | \omega \rangle}
{2\, m^4_{\rho}}\, [t_z (1) +  t_z (2) ] \,. \eqno (7a)
$$
Comparison with eqn.\ (6) shows that this mechanism provides a contribution
to $\gamma_s$:
$$
\gamma_s^{\rho \omega} = 
\frac{g_{\rho} g_{\omega} \langle \rho^0 | H | \omega \rangle }{2\, m^4_{\rho}}
\equiv c_{\rho \omega}(\epsilon \frac{m^2_{\pi}}{f^2_{\pi} \, \Lambda^2})
\, .  \eqno (7b)
$$
Using standard values for coupling constants and matrix elements measured in
$e^+ + e^- \rightarrow \pi^+ + \pi^-$ in storage rings for $s \sim m^2_{\omega}$
($g_{\rho} \sim 5.5, g_{\omega} \sim 15.2$, and $\langle \rho^0 | H | \omega
\rangle \sim -4500$ MeV$^2$ \cite{hm,msc}) we can calculate $\gamma_s^{\rho
\omega} \sim -0.54$ GeV$^{-2}$. Using $\Lambda \sim m_{\rho}$, we can solve 
eqn.\ (7b) for the dimensionless parameter $c_{\rho \omega}$ and find $c_{\rho
\omega} \cong -0.5$, a value that is not unnatural. The on-shell mixing-matrix
element ($\cong -0.8 (\epsilon \, m^2_{\pi})$), the $\rho NN$ coupling constant
\cite{ksfr} ($g_{\rho}/m_{\rho} \cong 0.66 / f_{\pi}$) and the $\omega NN$
coupling constant ($g_{\omega}/m_{\omega} \cong 1.8 /f_{\pi}$) are also natural.
Therefore we can conclude that the {\it assumption} of saturation of $\gamma_s$
by  on-shell $\rho - \omega$ mixing is consistent with naturalness. This has an
important consequence. The fact that the on-shell $\rho - \omega$ mixing model
does provide the necessary CSB in the $NN$ system implies that the latter can
more generally be explained by a chiral Lagrangian with natural parameters. Then
it is obvious that any mechanism or sum of mechanisms that produces a $\gamma_s$
of natural size (and correct sign) can potentially account for the observed CSB,
both in $NN$ scattering and in the $^3$He - $^3$H mass difference. For example,
a big off-shell suppression of $\rho - \omega$ mixing (as suggested, e.g., in
ref.\cite{ght}) would mean that we have to relax the saturation assumption so
that other processes (including those not involving resonances) restore
naturalness. This demonstrates the strength of the model-independent chiral
perturbation theory approach advocated here.

Other meson-mixing forces can also be calculated.  The $\pi - \eta$ mixing 
force \cite{msc,cp} has the form (and again neglecting $\mbox{\boldmath $q$}^2$
with respect to $m^2_\eta$): 
$$
V_{\pi \eta} (\mbox{\boldmath $q$}) = \frac{ g_A \overline{g}_{\eta} 
\langle \pi^0 | H | \eta \rangle }{2 f^2_{\pi}\, m^2_{\eta}} 
\left[\frac{\mbox{\boldmath $\sigma$}(1)  
\cdot \mbox{\boldmath $q \sigma$}(2) \cdot \mbox{\boldmath $q$}}
{(\mbox{\boldmath $q$}^2 + m^2_{\pi})} \right] (t_z(1) + t_z(2)) \, , 
\eqno (8a)
$$
where we have defined $g_{\eta}/M \equiv \overline{g}_{\eta} / f_{\pi}$ and have
used the GT relation. Comparing to eqn. (4) leads to a model-dependent
prediction for $\beta_1$:
$$
\beta_1^{\eta} = \frac{
\overline{g}_{\eta}}{m^2_{\eta}} \langle \pi^0 | H | \eta \rangle 
\equiv c_{\eta} (\frac{\epsilon \, m^2_{\pi}}{\Lambda^2}) \, , \eqno (8b)
$$
Using \cite{hm,msc,cp} $\overline{g}_{\eta} \sim 0.25$, $m_{\eta}$ = 550 MeV,
and $\langle \pi^0 | H | \eta \rangle \sim -4200$ MeV$^2 (\cong -0.7 (\epsilon
m^2_\pi ))$, we find $\beta_1^{\eta} \sim -3.5 \cdot 10^{-3}$. This is less than
the upper limit for $\beta_1$ that we found earlier. Using $\Lambda \sim$ 1 GeV,
we can solve eqn.\ (8b) for $c_{\eta}$ and find $c_{\eta} \cong -0.6$, so that
saturation of $\beta_1$ by on-shell mixing is also consistent with naturalness.
This force does not include the effect of $\pi -\eta^{\prime}$ mixing, which may
be nonnegligible. Studies of $\pi - \eta$ mixing suggest that it is not greatly
suppressed in a nucleus \cite{km}.

What simple mechanism could, likewise, produce a natural $\gamma_{\sigma}$ term
in eqn.\ (6)? If we restrict ourselves to single-meson exchanges in the static
limit without derivative coupling (which are simplest and perhaps dominant),
that meson must couple to one of the Dirac operators (1, $\gamma_5$,
$\gamma^{\mu}$, $\gamma_{5} \gamma^{\mu}, \sigma^{\mu \nu}$).  The matrix
$\gamma_5$ has no static limit, while $\gamma^{\mu}$ and 1 have a
spin-independent static limit. A tensor meson has a symmetric spin tensor, which
cannot couple to the antisymmetric $\sigma^{\mu \nu}$.  This leaves $\gamma_5
\gamma^{\mu}$, whose static limit is $\mbox{\boldmath$\sigma$}$.  Hence,
axial-vector $(1^+)$ meson exchange can generate $\gamma_{\sigma}$.  Candidates
\cite{pd} with appropriate isospins are $h_1 (1170) - b_1 (1235)$ and $a_1
(1260) - f_1 (1285)$.

We conclude that the isospin-violating mechanisms that have been used to explain
the $^3$He - $^3$H mass difference are of natural size, although the effect of
axial-vector-meson mixing has never been calculated (or considered heretofore).
In particular, as we have shown, there is no evidence for large isospin
violation in the $\pi NN$ coupling constants, one of which subsumes the effect
of $\pi - \eta$ mixing. If the $\rho - \omega$ mixing force is much smaller than
the dimensional estimate, one expects some other mechanism to restore the
naturalness of $\gamma_s$; otherwise, we would be facing a less than ideal
situation where a number of small effects would have to conspire coherently to
make up the difference. However, the more important conclusion transcends
models. We find that the observed magnitude of CSB by class (III) nuclear forces
results from the symmetries and naturalness of QCD. This is the power of chiral
perturbation theory: questions of specific mechanisms need neither be posed nor
answered.

{\Large {\bf Acknowledgments}}

This work was performed under the auspices of the United States Department of  
Energy. We would like to thank R. Timmermans of KVI and J. de Swart of
Nijmegen for helpful conversations, and R. Klomp of Nijmegen for generously
providing his preliminary np results.

\end{document}